# Boron vacancies in bulk h-BN created by high-energy He+ irradiation

P.G. Vilyuzhanina[1,2,3,4], S.V. Bolshedvorskii[1,4], R. Isayev[5,6], A.S. Doroshkevich[5,7], A.A. Tatarinova[5], V.V. Soshenko[1,4], I.S. Cojocaru[1,2], A.M. Kozodaev[1,2,3], S.M. Drofa[1,2,3,4], A. Chernyavskiy[1,2,3], N.I. Salangin[1,3], A.N. Smolyaninov[4], A.Y. Kuntsevich[1], A.V. Akimov[1,2,4]

[1]*P.N. Lebedev Physical Institute of the Russian Academy of Sciences, 119991 Moscow, Russia*

[2]*Russian Quantum Center, 143025 Moscow, Russia*

[3]*Moscow Institute of Physics and Technology (National Research University), 141701 Dolgoprudny, Russia*

[4]*LLC "Diamond Sensors", 121205 Moscow, Russia*

[5]*Joint Institute for Nuclear Research, 141980 Dubna, Russia*

[6] *IDDA Department of Nuclear Research, AZ1069 Baku, Azerbaijan*

[7]*Dubna State University, 141982 Dubna, Russia*

While color centers in diamond and other three-dimensional crystals are nowadays key elements of several types of sensors, color centers in two-dimensional materials are rapidly developing and promise high-tech applications. One of the interesting centers in 2D materials is the negatively charged boron vacancy in hexagonal boron nitride (h-BN), which has already shown some potential for magnetometry, but the reliable creation of boron vacancies remains challenging. Here, we demonstrate the fabrication of negatively charged boron vacancy color centers in bulk h-BN via implantation of helium ions at ~1 MeV, as confirmed by characteristic photoluminescence and optically detected magnetic resonance. The resonance has a width of 180 MHz and exhibits the expected Zeeman shift of the resonance lines. The depth of the color center along the c-axis of the h-BN crystal was measured and compared with predictions from Stopping and Range of Ions in Matter modeling. The experimental value and the modeled prediction are consistent within the reported uncertainties, as their 1σ intervals overlap, although the experiment showsa greater depth and a wider distribution.

## I. INTRODUCTION

Quantum technologies are rapidly advancing, requiring robust platforms for practical implementation [1]. While manipulating single atoms and ions necessitates complex and expensive setups, such as vacuum chambers and cryogenic cooling, color centers in wide-bandgap materials offer an alternative as quantum systems operable under ambient conditions. Color centers can be utilized as sources of single photons [2–4], as qubits [5,6] and as precise quantum sensors, with especially high sensitivity to magnetic fields, down to fractions of $\mathrm{pT}/\sqrt{\mathrm{Hz}}$ [7–12].

Currently, the research on color centers focuses on three-dimensional materials, primarily diamond, and in particular to the NV center [13]. However, the scientific community is looking for color centers in alternative and cheaper materials: silicon carbide, yttrium aluminum garnet, and zinc oxide [14].

Two-dimensional materials, particularly hexagonal boron nitride (h-BN), are of great interest. It is a dielectric with a band gap of 6 eV and lattice constants $a = 2.5\,\mathrm{\mathring{A}}$ and $c = 6.7\,\mathrm{\mathring{A}}$. h-BN can be easily combined with the other two-dimensional materials in stacked structures and devices [15–18]. Color centers in two-dimensional materials are more accessible than those in three-dimensional materials, which is very promising for local quantum sensing and magnetometry [19,20]. One of the best sensitivities reported in various publications devoted to $\mathrm{V_B^-}$ reaches $2.55\,\mu\mathrm{T}/\sqrt{\mathrm{Hz}}$ [20].

Deterministic fabrication of color centers with the desired concentration and localization in h-BN crystals requires serious efforts. Specifically, detection of $\mathrm{V_B^-}$ in pristine h-BN crystals has not yet been reported. Methods of color center creation can be divided into two broad categories: top-down and bottom-up. Bottom-up approaches are growth techniques; color centers are created through the addition of impurities during growth and variation of growth parameters. For example, a carbon source can be added during metal–organic vapor-phase epitaxy or molecular beam epitaxy of h-BN for color center formation [21].

Top-down approaches involve the creation of defects in the crystal, usually using various irradiation techniques: irradiation with laser pulses [22,23], electrons [24–28], ions with relatively low energies of ~10 keV [29–35], ions with high energies of ~ 1 MeV [19,20,36,37], and neutrons [38,39]. Annealing [40], chemical treatment, and mechanical damage induced by STM or AFM tips [41,42] can also be classified as top-down techniques for color center creation. Some of these top-down approaches are preferable for the creation of dense ensembles of color centers due to their large-scale implementation, namely annealing, chemical treatment, and irradiation with an unfocused particle beam [41]. Meanwhile, other methods, especially the tip method or irradiation with a focused beam of particles form color centers with high spatial resolution, which is highly desirable for single photon emission [41].

Despite the widespread use of irradiation methods, predicting the resulting damage to the sample and types of produced color centers is challenging. Several approaches can be used to evaluate the penetration depth of different ions into h-BN, namely molecular dynamics simulations [37,43,44] and the Stopping and Range of Ions in Matter (SRIM) software [45]. The latter models the interaction of charged particles with solids. It allows calculation of the ion or electron range in a material, energy loss, and the distributions of implanted particles and vacancies formed by these interactions [46].

It has been experimentally shown that SRIM modeling underestimates the depth of negatively charged boron vacancies created in h-BN by ion irradiation [35]. Possible reasons for this discrepancy are secondary ion processes and ion channeling [35,43], to avoid the latter some experimentalists tilt the h-BN sample during irradiation [31,38]. Backscattered ions and sputtered ions from the substrate should be considered for correct simulation of the defect formation [43].

Previous studies utilized irradiation of h-BN with ions with energies up to ~1 MeV, but they considered thin flakes [20,36,37] or hot-pressed BN and ceramics [19,37,47]. In this work, we irradiated bulk h-BN with $He^+$ particles at an energy of ~1 MeV and experimentally measured the photoluminescence depth profile along the *c*-axis of the h-BN crystal to determine the depth of the created color centers. We also performed optically detected magnetic resonance (ODMR) experiments and demonstrated the successful creation of $V_B^-$ color centers, exhibiting the Zeeman effect.

## II. THE SAMPLE

The bulk h-BN crystal, purchased from HQ Graphene, was placed on a copper plate and fixed with a TEM copper grid having a mesh of 400 lines per inch, corresponding to a distance of $63.5\,\mu m$ between the centers of neighboring holes, the so-called pitch. The sample was then irradiated with $He^+$ particles at an energy of $1.1\,MeV$ and a fluence of $5.7 \cdot 10^{15}\,ions/cm^2$ using the EG-5 electrostatic accelerator (I.M. Frank Laboratory of Neutron Physics, Dubna, Russia). A photograph of the non-irradiated crystal is shown in Figure 1(a), while a photograph of the irradiated crystal fixed on a copper plate with a TEM copper grid is shown in Figure 1(b).

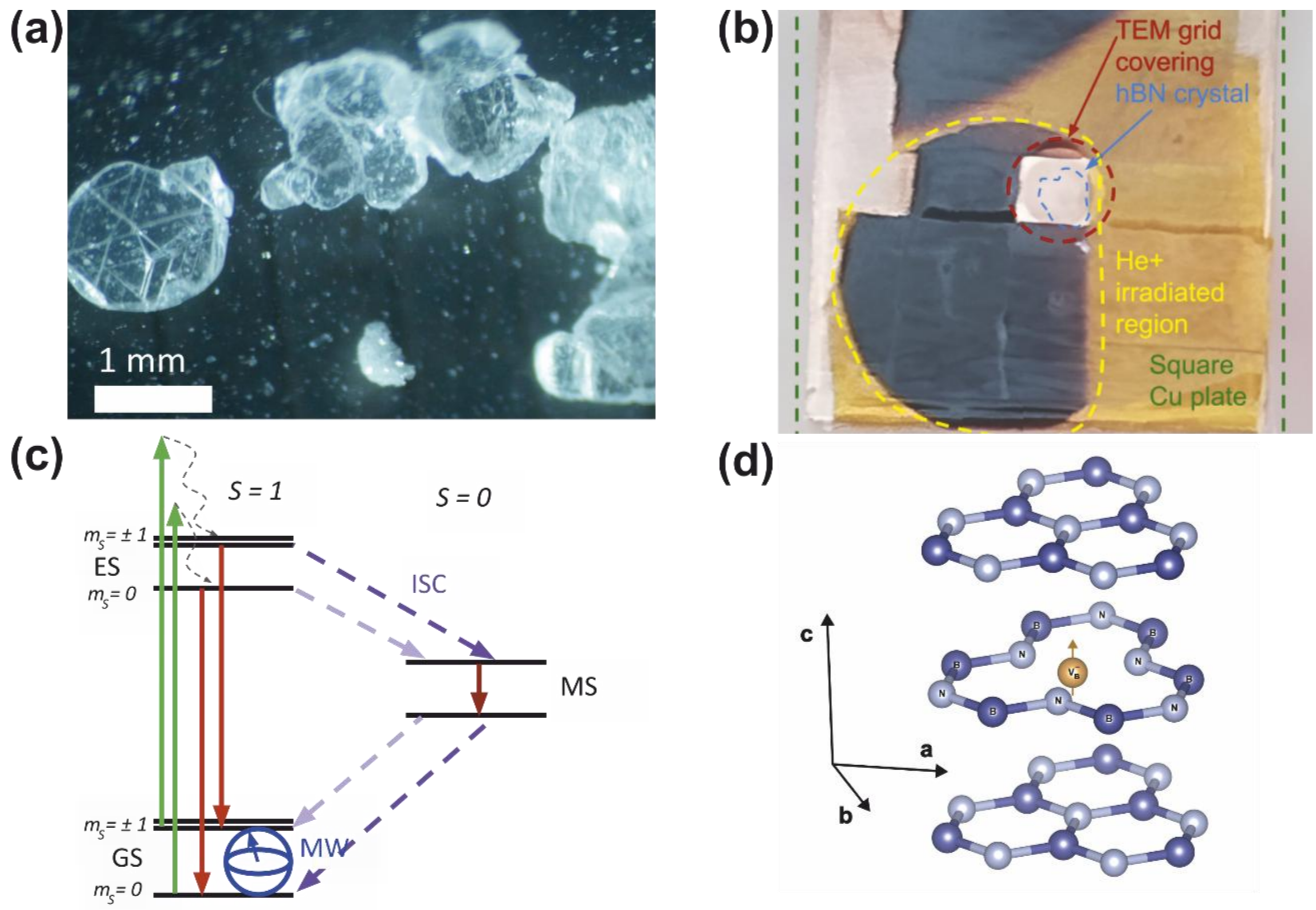

*Figure 1. a) photo of the non-irradiated crystals, the scale bar is 1 mm; b) photo of the irradiated sample c) simplified energy level structure of the $V_B^-$ color center and its transitions. Red color, solid lines – radiative transitions, purple dashed – non-radiative (ISC – intersystem crossing), green solid – optical pumping, blue solid – microwave (MW) driving. GS – ground state, ES – excited state, MS – metastable state [48]; d) schematic image of the h-BN crystal lattice with a boron vacancy (VB).*

## III. PHYSICAL SYSTEM

The negatively charged boron vacancy $\mathrm{V_B^-}$ is the most promising color center for quantum magnetometry in h-BN. The electron Hamiltonian describing $\mathrm{V_B^-}$ with electronic spin $S=1$ in the presence of an external magnetic field $\vec{B}$ is given by the following equation [41]:

$$H_e = D\left(S_z^2 - \frac{S(S+1)}{3}\right) + E\left(S_x^2 - S_y^2\right) + \gamma_e \vec{B}\vec{S} \tag{1}$$

where $D/h \approx 3.48\,\mathrm{GHz}$ is the zero-field splitting parameter, and $E/h \sim 50\,\mathrm{MHz}$ is the orthorhombic splitting of $\mathrm{V_B^-}$, $\gamma_e = 28\,\mathrm{GHz/T}$ is the gyromagnetic ratio, $h$ is the Planck constant, and $\vec{S}$ is the electron spin operator. The first two terms in (1) describe the electron spin–spin interaction, and the last term is the electron Zeeman effect.

In the presence of a magnetic field $\vec{B}$, the energy levels with spin projections $m_S = \pm 1$ experience a Zeeman shift. The frequencies $\nu_\pm$ of the quantum transitions $m_S = 0 \rightarrow m_S = \pm 1$ can be found from the following equation [41]:

$$\nu_\pm \approx \frac{D}{h} \pm \sqrt{\left(\frac{E}{h}\right)^2 + \left(\gamma_e B_z\right)^2} \tag{2}$$

where $B_z$ is the projection of the magnetic field onto the axis of the $V_B^-$, which is the out-of-plane axis of the two-dimensional h-BN (*c*-axis). Equation (2) is key to understanding the magnetic field sensing capability of $V_B^-$. By measuring the frequencies $\nu_\pm$ in the experiment, the projection of the magnetic field onto the $V_B^-$ color center axis $B_z$ can be calculated using a simple equation:

$$|B_z| = \frac{1}{\gamma_e} \sqrt{\left(\frac{\nu_+ - \nu_-}{2}\right)^2 - \left(\frac{E}{h}\right)^2} \tag{3}$$

The frequencies $\nu_\pm$ can be determined using the ODMR method. It allows to readout the spin of the system by the spin-dependent photoluminescence. Figure 1(c) shows a simplified diagram of the energy levels, including states with different electron spin projections [48]. Figure 1(d) presents a schematic image of the h-BN crystal lattice with a boron vacancy.

First, let us consider the case of a zero magnetic field $B = 0$. The sample is pumped optically with the laser. Due to laser pumping, the system undergoes a transition to an excited state (green arrows in Figure 1(c)). The transition to the ground state occurs with the emission of phonons and photoluminescence photons (red arrows in Figure 1(c)). When the frequency of the external microwave field (blue sphere in Figure 1(c)) coincides with the frequency of one of the transitions $m_S = 0 \rightarrow m_S = \pm 1$, resonant absorption occurs, driving the corresponding transition [49]. In the excited state, the system is more likely to relax to a metastable state if the spin projection is $m_S = \pm 1$ rather than $m_S = 0$, resulting in a decrease in photoluminescence intensity. When a resonant microwave field is applied, a photoluminescence decays. In other words, a negative ODMR contrast is observed.

When an external magnetic field is applied, the Zeeman effect shifts the energy levels, and consequently, the resonance frequencies. Thus, by detecting the photoluminescence intensity at various frequencies of the driving microwave field, it is possible to determine the magnetic field.

## IV. PHOTOLUMINESCENCE AND ODMR STUDIES

The irradiated h-BN sample was studied using a home-built confocal microscope equipped with a microwave module for ODMR measurements. A simplified schematic of the setup is shown in Figure 2. The green line in Figure 2 represents the 532 nm laser, which is focused onto the sample using a high-numerical-aperture $NA = 0.95$ objective. The red line represents the photoluminescence from the sample, which is collected through the same objective and then detected by avalanche photodiodes. Galvo mirrors are used to scan the sample in the lateral dimensions. The setup contains two photodiodes (two channels): only one channel was used in this experiment; the second one is shown to accurately represent the optical path, including the BP145B2 pellicle beamsplitter (Thorlabs), labeled BS in Figure 2.

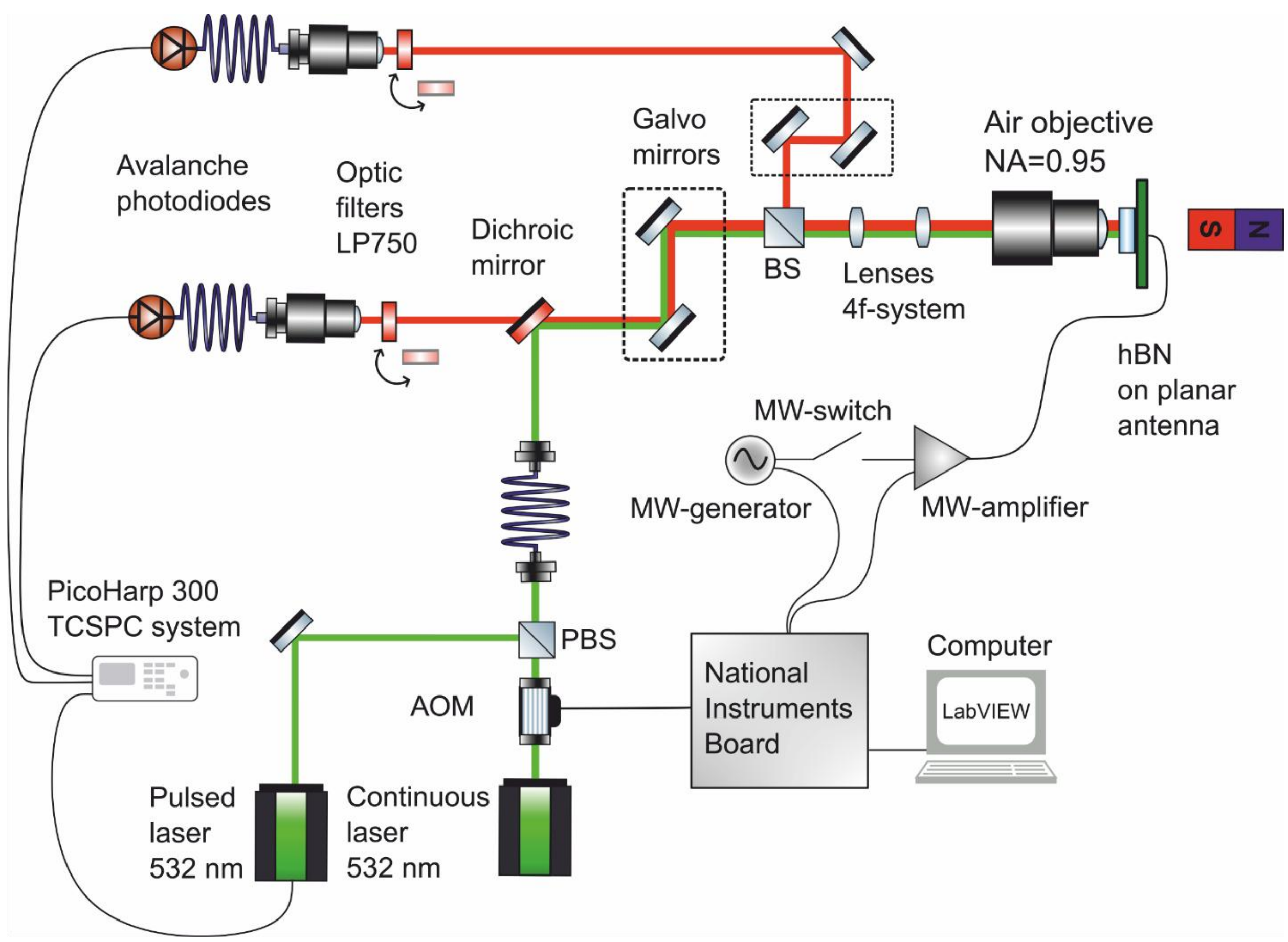

*Figure 2. Simplified schematic of the home-built confocal microscope setup.*

For ODMR studies and pulse sequence generation, the required sequence is programmed in LabVIEW. Subsequently, a National Instruments board BNC-2121 with NI-DAQmx driver, a microwave switch, and an acousto-optic modulator are used to switch the microwave and laser radiation in the correct order. A permanent magnet on a moving stage is used to cause the Zeeman effect.

The ion penetration depth into the crystal was studied by moving the objective along the c-axis of the h-BN crystal using a piezo actuator (E-665 LVPZT Amplifier).

First, photoluminescence measurements were performed to confirm the formation of color centers and $V_B^-$ centers specifically by irradiation. Confocal scans of the sample are shown in Figure 3(a,b). The photoluminescent regions exhibit a signal of up to $2 \cdot 10^6$ counts / sec. We measured the dependence of the photoluminescence intensity on laser power up to 3.6 mW and observed no saturation (for details, see Supplementary Materials, Figure S1(a)).

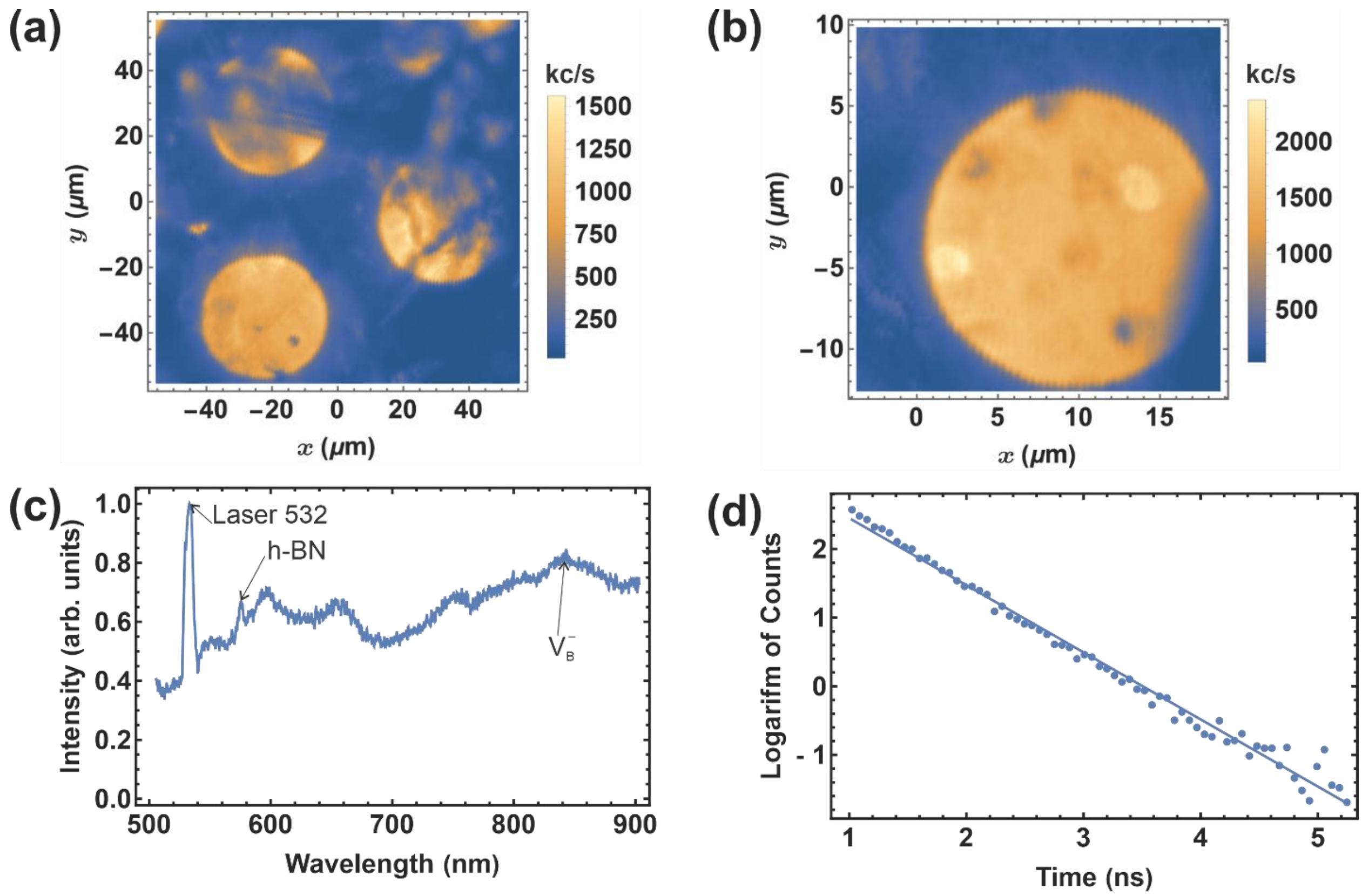

*Figure 3. Confocal scans of the irradiated bulk h-BN crystal. a) overview scan; b) zoomed-in region; c) photoluminescence spectrum; d) intensity of a color center versus time on a logarithmic scale.*

In the confocal scans, circular patterns of the TEM copper grid are clearly observed. The area of the sample not covered by the copper grid shows a bright, relatively spatially homogeneous signal due to formed color centers.

Next, we measured the photoluminescence spectrum of the irradiated region using an M266 monochromator (SOLAR LS) and a Dhyana 400D camera with high quantum efficiency, the resulting spectrum is shown in Figure 3(c). The observed peak in the photoluminescence spectrum in the 800–850 nm range is typical of $V_B^-$ color centers and agrees with theoretical predictions [50] and other experiments [28,33,51]. The sharp line at 532 nm corresponds to the laser line; the line at ~570 nm is the h-BN Raman mode, confirming that the crystal structure was preserved after irradiation. The photoluminescence in the 600–650 nm range may be related to $N_B V_N$ color centers [52–54].

For the optical spectrum measurement, the optical fiber of the confocal microscope (see Figure 2) was switched from the avalanche photodiode to the monochromator input. Note that the photoluminescence spectrum is modulated by the pellicle beamsplitter BP145B2 (Thorlabs); its transmission and reflectance coefficients vary periodically by approximately 20–30% over a wavelength interval of about 100 nm [55].

We also performed lifetime measurements using a pulsed laser and a Time-Correlated Single Photon Counting system (PicoHarp 330). The experimental results are shown on a logarithmic scale in Figure 3(d). The experimental data were fitted with a linear function, and the estimated lifetime is $\tau = 1.02 \pm 0.03$ ns, which is in reasonable agreement with literature value $\tau = 1.2$ ns [56].

The irradiated photoluminescent regions of the sample were studied using the pulsed ODMR technique. The optimal microwave pulse length was selected from the experimental dependence of the contrast on the pulse length (for details, see Supplementary Materials, Figure S2(a)). However, it is worth noting that we did not observe Rabi oscillations, probably due to the low microwave power used in our experiments. Using the optimal MW pulse length, we also assessed the effective electron spin relaxation time $T_1^{eff} = 16 \pm 11\,\mu s$, (for details, see Supplementary Materials, Figure S2(b)).

We applied three different external magnetic fields $B$ along the c-axis of the h-BN crystal to observe the Zeeman shift of the resonance lines. The resulting ODMR spectra, fitted with a sum of two Lorentzian functions, are shown in Figure 4(a). ODMR and photoluminescence results confirm the creation of $V_B^-$ in the bulk h-BN crystal sample upon irradiation.

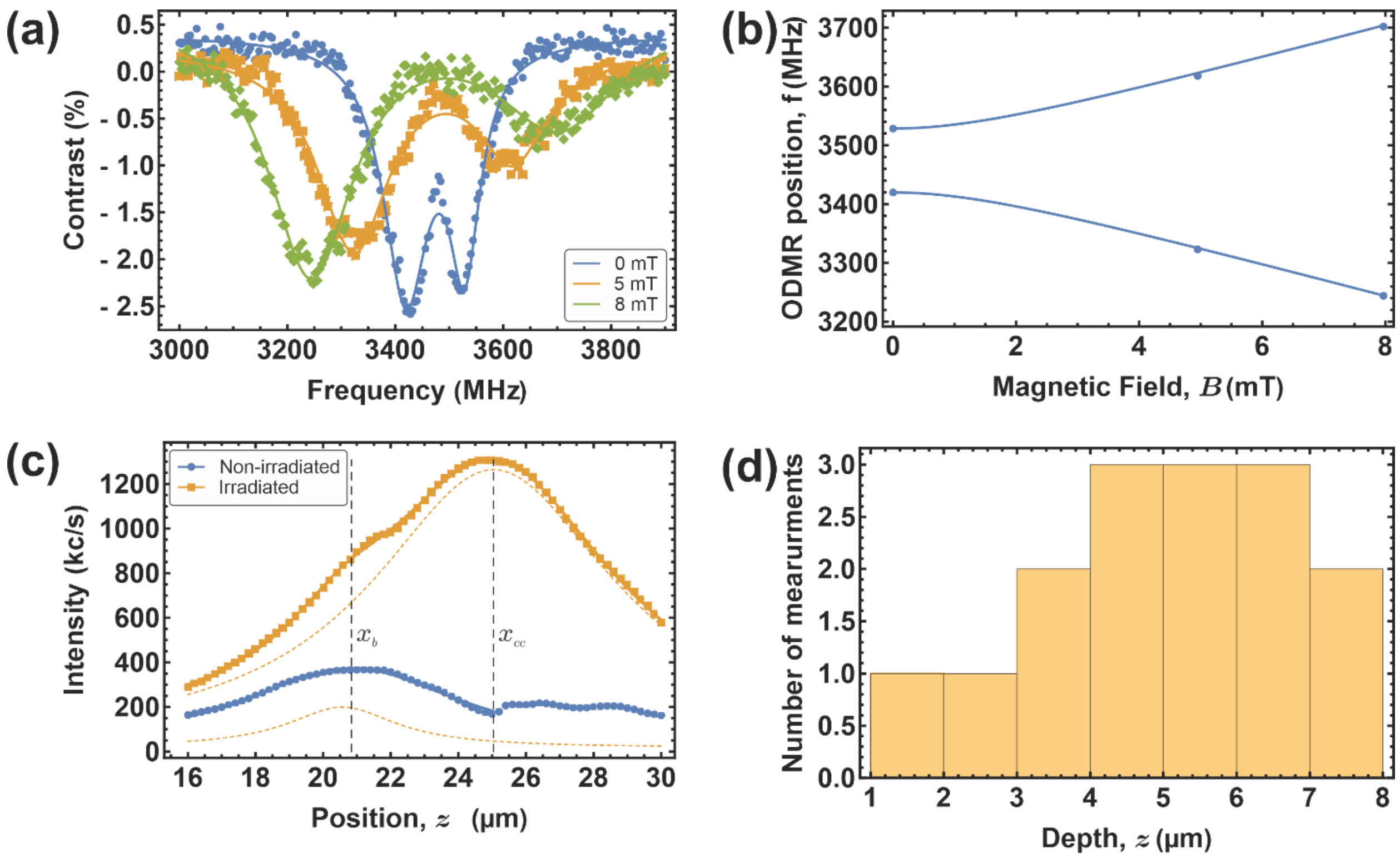

*Figure 4. a) experimental ODMR spectra of the irradiated h-BN crystal for different external magnetic fields; b) fitted resonance frequencies from the ODMR spectra plotted versus the estimated magnetic field; c) dependence of the signal from non-irradiated (reflection) and irradiated (photoluminescence and reflection) areas on the objective position during scanning along the c-axis of the h-BN crystal. Orange dashed lines represent the Lorentzian fit, vertical black dashed lines represent the crystal boundary $x_b$ and the depth of the color centers $x_{CC}$; d) histogram of the estimated experimental depths of color centers.*

The values of the zero-field splitting parameter $D/h = 3474.1 \pm 1.2$ MHz and the orthorhombic splitting $E/h = 54.4 \pm 1.2$ MHz were estimated by fitting the experimental ODMR spectra in the absence of an external magnetic field $B = 0$. The obtained values agree with previous results [56] for the $V_B^-$ color center.

The values of the external magnetic field estimated from (3) and from the fitted data are $B_1 = 4.95 \pm 0.10$ mT and $B_2 \approx 7.97 \pm 0.13$ mT. The fitted resonance frequencies are plotted versus the estimated magnetic field in Figure 4(b). The spectra show a clearly observable shift of the spectral lines, as expected from the Zeeman effect. The asymmetry of the spectra may be attributed to the higher efficiency of the microwave antenna at lower frequencies.

The sensitivity to the magnetic field $\eta_B$ was calculated from the fitted continuous-wave spectrum (for details, see Supplementary Materials, Figure S3) using the following equation [20]:

$$\eta_B = P_F \cdot \frac{1}{\gamma_e} \cdot \frac{\Delta\nu}{C\sqrt{R}} \ , \tag{4}$$

where $\gamma_e = 28\ \mathrm{GHz/T}$ is the gyromagnetic ratio mentioned above, and $P_F$ is a line-shape parameter given by $P_F = \frac{4}{3\sqrt{3}} \approx 0.77$ for a Lorentzian function [57]. $R \approx 2.17 \cdot 10^6$ photons per second is the photon count rate. The full width at half maximum (FWHM) $\Delta\nu \approx 180$ MHz and contrast $C \approx 1\%$ were obtained from the fitted experimental data at $B \approx 8 mT$ and $\Delta\nu \approx 51\ MHz$ and contrast $C \approx 2.6\%$ at zero magnetic field correspondingly. The estimated sensitivities $\eta_B = 340 \frac{\mu T}{\sqrt{Hz}}$ and $\eta_B = 37 \frac{\mu T}{\sqrt{Hz}}$ for non-zero and zero magnetic field do not reach the previously reported value of $2.5\ \mu\mathrm{T}/\sqrt{\mathrm{Hz}}$ for zero magnetic field [20]. However, it is comparable to the sensitivity of $200\ \mu\mathrm{T}/\sqrt{\mathrm{Hz}}$ reported in [19]. As mentioned earlier, the contrast in our experiment can be improved by increasing the MW power, thereby enhancing the magnetic-field sensitivity.

## V. EXPERIMENTAL DEPTH SCAN

We determined the axial resolution of our setup to be about 5 μm. For this purpose, we scanned a single NV center in diamond in the same way as for the h-BN depth scan (for details, see Supplementary Materials, Figure S4). Then we performed depth scan measurements, representative depth scans of the irradiated and non-irradiated h-BN regions are shown in Figure 4(c). The intensity values and errors shown in Figure 4(c) were calculated from the time trace measured while the piezo actuator was fixed at a given position. The error of the piezo actuator corresponds to its instrumental error.

The signal from the non-irradiated area was fitted with a single Lorentzian function, while the signal from the irradiated area was fitted with a sum of Lorentzian functions. This functional dependence was chosen for fitting because the intensity at the center of a Gaussian beam as a function of the longitudinal coordinate follows a Lorentzian function.

The crystal boundary depth $x_b$ was assumed to correspond to the maximum reflection signal and to coincide with the center of the Lorentzian function fit of the non-irradiated area scan. The color centers depth $x_{CC}$ was assumed to correspond to the center of the second Lorentzian function in the sum from the fit of the irradiated area scan, while the first corresponds to the boundary of the crystal.

Note that the depth can be estimated without a depth scan of the non-irradiated area by considering only the fit of the irradiated area depth scan, but we used the depth scan of the non-irradiated area to reduce the uncertainty.

The actual depths were estimated by subtracting the boundary position from the photoluminescence peak position and multiplying the result by the refractive index:

$$d_{exp} = n_{\perp} \cdot \left( x_{CC} - x_b \right) , \tag{5}$$

where $n_{\perp} = 1.8$ is the refractive index of h-BN along the *c*-axis [58]. The in-plane h-BN refractive index is slightly higher, $n_{\parallel} = 2.1$, considering the light path inside the sample, the depth may be underestimated. The depth values range from 2 to 7 µm, the median depth is $d_{exp} = 5.2$ µm, and the standard deviation is $\sigma_{exp} = 1.7$ µm, the resulting distribution is presented in Figure 4(d).

## VI. STOPPING AND RANGE OF IONS IN MATTER MODELING

The SRIM modeling [46] of $He^+$ ions with energy $E = 1.1$ MeV in boron nitride with a density of $2.1 \mathrm{g/cm^3}$ was performed. We tried different displacement threshold energies [24,37,44] for boron $T_d^B$ and nitrogen atoms $T_d^N$ and found no differences in the median depth of the vacancy distribution. The results of SRIM modeling for $10^4$ ions and displacement threshold energies of $T_d^B = 40.5\mathrm{eV}$ and $T_d^N = 41.6\mathrm{eV}$ [44] are shown in Figure 5. The $He^+$ ion range distribution is presented in Figure 5(a), and the induced vacancies distribution is presented in Figure 5(b). The simulated median depth of the created vacancies is $3.1\mu\mathrm{m}$, and the standard deviation is $0.6\mu\mathrm{m}$.

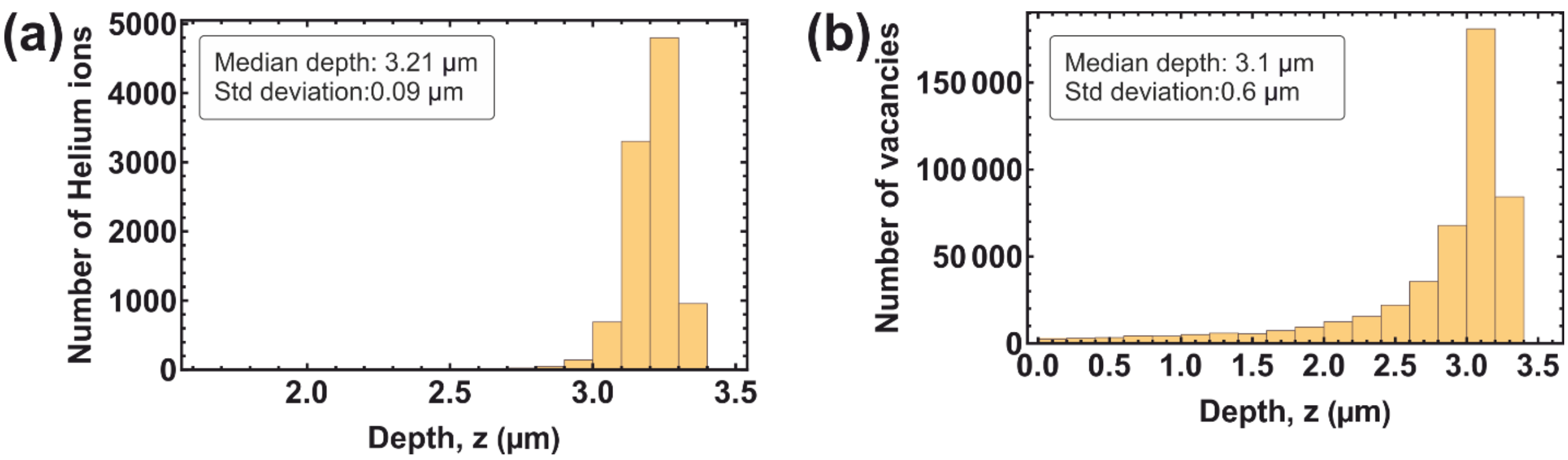

*Figure 5. Results of SRIM modeling of h-BN irradiated with 1.1 MeV $He^+$ particles. (a) ion range distribution; (b) depth distribution of irradiation-induced vacancies.*

Overall, the experimental depth is $d_{exp} = 5.2\mu\mathrm{m}$, with a standard deviation of $\sigma_{exp} = 1.7\mu\mathrm{m}$, while the depth modeled with SRIM is $d_{model} = 3.1\mu\mathrm{m}$ with a standard deviation of $\sigma_{model} = 0.6\mu\mathrm{m}$. We chose the median rather than the mean for comparison because it is more robust to the large variability observed in our measurements and to the asymmetry observed in the modeling. The values agree within the uncertainties, although the experimental measurements tend to indicate a greater depth. Possible reasons for that include channeling of $He^+$ ions, secondary ion processes [35], the unaccounted for two-dimensional crystal structure of h-BN, and its layered nature and inhomogeneity of the h-BN crystal.

High energy of the $He^+$ ions and the use of a thick crystal allowed rather deep implantation, which was deep enough to optically verify the SRIM modeling prediction of the depth within experimental uncertainties. Furthermore, the clean structure and large size of the crystal allowed us to avoid any substrate influence, thus forming color centers free from parasitic impurities. This irradiation method enables scalable, homogeneous, and robust boron vacancies for magnetometry.

Concerning further development, a desired spatial pattern of vacancies can be fabricated using a lithographically pre-patterned shadow mask to obtain spatial resolution. Another direction is the controllable exfoliation of top layers of the irradiated h-BN crystals. This technology should be developed on the basis of the known h-BN exfoliation methods [59] , and possibly include either hot pickup [60] or metal layer-assisted exfoliation [61] of the top layers. Thus, obtained thin flakes with different boron vacancy concentrations should demonstrate lower disorder and appear to be more sensitive to magnetic fields.

## VII. CONCLUSION

In this work, the first experimental irradiation of bulk h-BN with $He^+$ particles at an energy of ~1 MeV was performed. The successful scalable, homogeneous and robust fabrication of $V_B^-$ color centers by irradiation was demonstrated. They exhibit characteristic photoluminescence in the $800-900$ nm range and ODMR spectra, with experimentally estimated zero-field splitting parameter $D/h = 3474.1 \pm 1.2\,\text{MHz}$ and orthorhombic splitting $E/h = 54.4 \pm 1.2\,\text{MHz}$, in excellent agreement with previous results. ODMR measurements were performed under different static external magnetic fields, clearly showing the expected Zeeman shift of the resonance lines. The assessed sensitivity to the permanent magnetic field is $\eta_B = 340\,\mu\text{T}/\sqrt{\text{Hz}}$, which can be enhanced by increasing the power of the microwave field. The depth of the color centers was experimentally measured by detection of a photoluminescence along the $c$-axis of the h-BN crystal. The experimental depth $d_{exp} = 5.2 \pm 1.7\,\mu\text{m}$ and the modeled depth $d_{model} = 3.1 \pm 0.6\,\mu\text{m}$ overlap within the range of their standard deviations. $He^+$ irradiation method allows you to obtain a large number of homogeneously distributed vacancies in an area limited only by the diameter of the particle beam and size of the crystal. These results pave the way to controlled and scalable fabrication of $V_B^-$ color centers in h-BN and development of on-chip magnetometry with those.

# Acknowledgements

We thank our colleague M.V. Pugachev for his continuous technological support and fruitful discussions throughout this research. Sample fabrication was partially performed at P.N. Lebedev Physical Institute Shared Facility Center. Irradiation of the sample was performed in the I.M. Frank Laboratory of Neutron Physics (Dubna, Russia), this part of research was supported via cooperation projects with Serbia (No. 109 2026 item 9, No. 676 2025 items 7, 21, 26), Belarus No. 302 2026 item 18), Kazakhstan (No. 111 2026 item 13), and the Arab Republic of Egypt (No. 127 2026 item 8).